\documentclass[twocolumn,aps,prd,longbibliography, replace]{revtex4-1}
\usepackage{graphicx,bm,xcolor, bbold}
\usepackage[normalem]{ulem}
\usepackage{hyperref}
\hypersetup{colorlinks, linkcolor={blue},citecolor={blue},urlcolor={blue}}
\usepackage{graphicx,graphics}
\usepackage{dcolumn}
\usepackage{amsmath,amssymb,amsfonts}
\usepackage{latexsym,verbatim}
\usepackage{bm}
\usepackage[toc,page]{appendix}
\usepackage{color}
\usepackage[normalem]{ulem}
\usepackage{tabularx}
\graphicspath{fig}

\def\be{\begin{eqnarray}}
\def\ee{\end{eqnarray}}

\def\<{\left\langle}
\def\>{\right\rangle}

\newcommand{\beq}{\begin{equation}}
\newcommand{\eeq}{\end{equation}}

\newcommand\redsout{\bgroup\markoverwith{\textcolor{red}{\rule[0.4ex]{2pt}{2pt}}}\ULon}

\begin{document}
\title{electrical conductivity of the Quark-Gluon Plasma in the presence of strong magnetic fields}

\author{Giorgio~Almirante}
\email{giorgio.almirante@ijclab.in2p3.fr}
\affiliation{Université Paris-Saclay, CNRS/IN2P3, IJCLab, 91405 Orsay, France}
\author{Nikita~Astrakhantsev}
\email[]{nikita.astrakhantsev@physik.uzh.ch}
\affiliation{Physik-Institut, Universit{\" a}t Z{\" u}rich, Winterthurerstrasse 190, CH-8057 Z{\" u}rich, Switzerland}

\author{V.\,V.~Braguta}
\email[]{vvbraguta@theor.jinr.ru}
\affiliation{Bogoliubov Laboratory of Theoretical Physics, Joint Institute for Nuclear Research, Dubna, 141980 Russia}

%
\author{Massimo D'Elia}
\email{massimo.delia@unipi.it }
\affiliation{Dipartimento di Fisica dell'Universit\`a di Pisa, Largo B.~Pontecorvo 3, I-56127 Pisa, Italy}
\affiliation{INFN, Sezione di Pisa, Largo B.~Pontecorvo 3, I-56127 Pisa, Italy}
\author{Lorenzo\,Maio}
\email{lorenzo.maio@cpt.univ-mrs.fr}
\affiliation{Dipartimento di Fisica dell'Universit\`a di Pisa, Largo B.~Pontecorvo 3, I-56127 Pisa, Italy}
\affiliation{INFN, Sezione di Pisa, Largo B.~Pontecorvo 3, I-56127 Pisa, Italy}
\affiliation{Aix Marseille Univ, Universit\'{e} de Toulon, CNRS, CPT, Marseille, France}
\author{Manuel\,Naviglio}
\email{manuel.naviglio@sns.it}
\affiliation{Dipartimento di Fisica dell'Universit\`a di Pisa, Largo B.~Pontecorvo 3, I-56127 Pisa, Italy}
\affiliation{INFN, Sezione di Pisa, Largo B.~Pontecorvo 3, I-56127 Pisa, Italy}
\affiliation{Scuola Normale Superiore, Piazza dei Cavalieri 7, I-56126 Pisa, Italy}
\author{Francesco Sanfilippo}
\email[]{francesco.sanfilippo@roma3.infn.it}
\affiliation{INFN, Sezione di Roma Tre, Via della Vasca Navale 84, I-00146 Rome, Italy}

\author{Anton Trunin}
\email{amtrnn@gmail.com}

\begin{abstract}
{We compute the electrical conductivity of the strongly interacting medium in the presence of strong magnetic background fields, $eB=4,9~GeV^2$, and for different values of the temperature, both in the confined and in the deconfined Quark-Gluon Plasma (QGP) phase. The conductivity is obtained from the Euclidean lattice time correlator of the electrical current, computed on gauge configurations sampled from Monte-Carlo simulations of an improved staggered discretization of $N_f = 2+1$ QCD. We perform the inverse Laplace transform of the correlator adopting a recently-proposed version of the standard Backus--Gilbert procedure for the inversion. The results obtained in the QGP phase show a sizable enhancement of the conductivity in the direction parallel to the magnetic field, as well as a suppression in the direction orthogonal to it. Such enhancement could be attributed to the manifestation of the Chiral Magnetic Effect (CME): following this guess, we extract the behaviour of the relaxation time of this process, extrapolate it to the continuum limit and compare it to previous results, finding it lower than expected in the explored range of temperatures.}
\end{abstract}
\maketitle

\section{Introduction}
The presence of background magnetic fields affects the properties of strongly interacting matter, giving rise to several effects, that are interesting from both a theoretical and a phenomenological point of view. On the theoretical side, the effects of magnetic fields on the confining potential, on the chiral symmetry breaking and on the QCD phase diagram have been widely studied in the last few years~\cite{DElia:2021yvk,DElia:2021tfb,Cardinali:2022sjy,DElia:2022gam,DElia:2023jbe,Bali:2011qj,Bali:2012zg,Endrodi:2015oba,Endrodi:2023wwf}. From a phenomenological point of view, studying QCD in such extreme conditions could be useful in different fields. In the context of Heavy Ion Collisions (HIC) experiments, in particular for non-central collisions, intense magnetic fields are generated due to the charged particles moving at relativistic speeds towards each other with a non-zero impact parameter~\cite{Skokov:2009qp,Holliday:2016lbx,Tuchin:2013ie,Deng:2012pc}. On cosmological scales, extremely strong magnetic fields are expected to be generated in the Early Universe, during the cosmological electroweak phase transition \cite{cosmo}, having possibly had important consequences on the subsequent Universe evolution~\cite{Grasso:2000wj}. At astrophysical scales, magnetic fields of intensities up to $eB=10$~MeV$^2$ are expected to be found on the surface of Magnetars~\cite{Duncan:1992hi}. Being able to model the evolution of the QGP is crucial to make progress in all of these contexts, hence the importance of understanding the QGP behavior in strong background magnetic fields. 

The study of the influence of the magnetic field on transport coefficients, such as the electric conductivity, is of particular interest in the context of HIC: apart
from being relevant to the prediction of the evolution of QGP (produced during the scattering) and of the magnetic field itself, it plays a crucial role in the search for one of the most
long-hunted phenomena predicted for QGP in a magnetic background: the Chiral Magnetic Effect, an anomaly based phenomenon induced by the magnetic field in systems with relativistic fermions (see Ref.~\cite{Kharzeev:2024zzm} for a very recent review).

In the last years, a number of lattice studies were conducted on this phenomenon, through different techniques. For example, via the measurement of the electromagnetic current in an chirally imbalanced background~\cite{Yamamoto:2011gk,Yamamoto:2011ks}, see however also recent developments reported in Ref.~\cite{Brandt:2024wlw}. Other techniques include the definition of an Euclidean correlator of electric current and axial charge~\cite{Buividovich:2024bmu} through which it is possible to study the out-of-equilibrium effects of CME, or using the Kubo forumalae relating, in the framework of the linear-response theory, the spectral density to the transport coefficients~\cite{Amato:2013naa,Aarts:2014nba,Brandt:2015aqk,Ding:2016hua,Astrakhantsev:2019zkr}. 
In particular, in Ref.~\cite{Astrakhantsev:2019zkr} some of us studied the QGP transport properties in magnetic field and temperature ranges of interest for HIC experiments, respectively $0.3~{\rm GeV}^2\lesssim eB \lesssim 2.5~{\rm GeV}^2$ and $T=200$ and 250~MeV. The most interesting result was the observation of a strong enhancement of the electrical conductivity in the direction parallel to the magnetic field, accompanied by a moderate reduction in the direction transverse to it. The former could be interpreted as a possible manifestation of the Chiral Magnetic Effect (CME).

In this work we extend the results to two values of the magnetic field which can be of cosmological interest, namely $eB=4$ and 9~GeV$^2$, studying the QGP transport properties for different values of the temperature. This system is interesting on a theoretical level since it allows to gain insight into the QCD phase diagram: indeed, the transition has been proved to move down to substantially low temperature, well below 100~MeV, for such strong fields, and to turn
into strong first order for $eB= $~9~GeV$^2$. Therefore, this study can serve at the same time to give 
a better characterization of the two phases in terms of transport coefficients, and to study QGP properties, in particular those possibly 
related to CME, over an extended temperature range.

The present study is based on numerical Monte Carlo (MC) simulations of lattice QCD using the stout-smeared improved staggered fermion discretization to simulate 2+1 flavors of dynamical quarks. The paper is organized as follows: in Section~\ref{sec:CME} we summarize the main theoretical aspects
of the Chiral Magnetic Effect and of its possible manifestation in the electrical conductivity of the QGP; Section~\ref{sec:setup} provides details
on our lattice discretization and simulations algorithms, while in Section~\ref{sec:BG} we illustrate the method used to extract the electric conductivity
from Euclidean temporal correlators of the electric current; results are reported in Section~\ref{sec:results}, then we discuss their implications
and provide our conclusions in Section~\ref{sec:discussion}.

\section{CME and electrical conductivity}
\label{sec:CME}
The CME is a phenomenon emerging due to the anomalous breaking of the chiral symmetry, which can be realized in systems with relativistic fermionic degrees of freedom \cite{K1,K2}. The CME consists in the generation of a non-dissipative electric current along the external magnetic field in systems with a net imbalance in the number of right-handed and left-handed fermions, i.e., in systems with non zero chiral density $\rho_5$.
This mechanism can be described as follows: the presence of a magnetic field, $\vec B$, aligns the spins of the positive (negative) fermions in the direction parallel (anti-parallel) to $\vec B$. Then, a positive right-handed fermion, moves in the direction of the magnetic field, and in the opposite direction if left-handed; the contrary holds for negative fermions. Thus, when $\rho_5$ is finite, a net electric current is generated in the direction of the magnetic field. This current can be derived in various ways~\cite{Fukushima:2008xe}, and, in terms of the chiral chemical potential, $\mu_5$, it reads
\begin{equation}\label{eq:cme_curr}
    \vec{J}^{\,\,CME}= C_{em}\, \frac{N_c}{2\pi^2}\, \mu_5\, \vec{B},
\end{equation}
where $C_{em}=e^2\sum_fq_f^2$, $N_c$ is the number of colors, and $e$ and $q_f$ are, respectively, the positron electric charge and the $f$-flavored quark electric charge in units of $e$.

Eq.~\eqref{eq:cme_curr} can be manipulated in order to compute the electrical conductivity, $\sigma_{ij}$, that is defined as
\begin{equation}\label{eq:em_cond}
    J_i=\sum_j\sigma_{ij}E_j.
\end{equation}
Let us recall that a non-zero $\rho_5$ emerges due to the axial anomaly (both in high energy systems~\cite{K3} and condensed matter ones~\cite{lowCME}), that, in the presence of external electric and magnetic fields, gives rise to a finite divergence of the axial current, $J_A$, which reads
\begin{equation}\label{eq:axial_ano}
    \partial_{\mu}J_A^{\mu} =
    C_{em}\, \frac{N_c}{2\pi^2} \vec{E}\cdot\vec{B},
\end{equation}
where $\vec E$ is used to denote the external electric field. Considering the presence of chirality-changing scattering events, which, in the absence of chiral anomaly, suppress $\rho_5$ with a relaxation time $\tau$, one can obtain from Eq.~\eqref{eq:axial_ano} the chiral density generation rate~\cite{K4} as
\begin{equation}\label{eq:chirden_rate}
    \frac{d\rho_5}{dt} =
    C_{em} \frac{N_c}{2\pi^2} \vec{E}\cdot\vec{B} -
    \frac{\rho_5}{\tau}.
\end{equation}
This equation has stationary solution
\begin{equation}\label{eq:stat_sol}
    \rho_5 =
    C_{em} \frac{N_c}{2\pi^2} \vec{E}\cdot\vec{B}
    \hspace{1mm} \tau,
\end{equation}
which describes the balance between chiral-anomaly production rate and chirality relaxation processes. Finally, the equation of state $\rho_5=\rho_5(\mu_5)$ allows to parameterize the density $\rho_5$ in terms of the chiral chemical potential $\mu_5$, and hence to use the linear response theory by considering the electric field as a perturbation: the resulting chiral chemical potential is small and the equation of state reads
\begin{equation} \label{eq:eos}
    \rho_5= \mu_5 \chi(B,T) + O(\mu_5^3),
\end{equation}
where the linear response function $\chi(B,T)$ only depends on the background magnetic field intensity and the temperature. Thus, we can use Eqs.~\eqref{eq:stat_sol} and~\eqref{eq:eos} to substitute $\mu_5$ in Eq.~\eqref{eq:cme_curr}, obtaining the following expression for the conductivity
\begin{equation} \label{eq:sigma_CME}
    \sigma_{zz}^{CME}=\frac{N_c^2\, C_{em}^2}{4\pi^4} \frac{B_z^2}{\chi(B,T)}\tau,
\end{equation}
where we only consider the $i=j=z$ entry of the conductivity, as, for convenience, we set $B_x=B_y=0$. As we are interested in the large magnetic field limit, i.e., $\sqrt{eB} \gg T$, the chiral density equation of state, Eq.~\eqref{eq:eos}, can be considered as governed by the lowest Landau level degeneracy,  which, in the non interacting approximation, implies
\begin{equation}\label{eq:highB_approx}
    \chi(B,T)=N_c\sum_f |q_f| \frac{eB_z}{2\pi^2}.
\end{equation}
Then, Eq.~\eqref{eq:sigma_CME} can be written as
\begin{equation}\label{eq:approx_sigmaCME}
    \sigma^{CME}_{L} =
    \frac{N_c\,C_{em}^2}{2\pi^2\,e^2}
    \frac{eB}{\sum_f |q_f|}\tau,
\end{equation}
where we dropped all the $z$ subscripts, and the $L$ means that the equation is only valid in the direction longitudinal to $\vec{B}$.
In this paper, we study the dependence of the conductivity on the temperature in a strong background magnetic field. Eq.~\eqref{eq:approx_sigmaCME} shows that, in the large magnetic field limit, such dependence can be attributed to the temperature behaviour of the relaxation time $\tau$, which will be discussed in Section~\ref{sec:discussion}. 

\section{The Lattice Setup}
\label{sec:setup}

\begin{table*}[t]
    \centering
    \begin{tabular}{c|c|c|c|c|c|c}
        \hline $eB$[GeV$^2$]&$b$&$a$[fm]&$\beta$&$a m_s$&$N_s$&$(aT)^{-1}$  \\ \hline
        &&0.0572&4.140(6)&0.0224&48&12, 14, 16, 18, 20, 22, 24, 30, 32, 34, 36, 38, 40\\
        9&93&0.0858(2)&3.918&0.0343&32&12, 14, 16, 18, 20, 22, 24, 26\\
        &&0.1144(3)&3.787&0.0457&24&12, 14, 16, 18, 20, 22\\ \hline
        &&0.0572&4.140(6)&0.0224&48&12, 14, 16, 18, 20, 30, 32, 34, 36, 38, 40\\
        4&41&0.0858(2)&3.918&0.0343&32&12, 14, 16, 18, 20, 22, 24, 26\\
        &&0.1144(3)&3.787&0.0457&24&12,14,16,18,20,22\\ \hline
    \end{tabular}
    \caption{Choices of lattice spacing, volumes and magnetic fields for simulations. The physical line is based on \cite{LCP1,LCP2} and it corresponds to the physical value of the pion mass. The strange-to-light mass ratio is $m_s/m_l=28.15$. The systematic error on $a$ is about $2\%$.}
    \label{parameters}
\end{table*}
Results reported in this work are based on lattice computations of current--current correlators performed on part of the ensembles generated to conduct the analyses reported in Ref.~\cite{DElia:2021yvk}. We briefly revise here the adopted lattice setup.

Configurations have been generated by means of Rational Hybrid Monte Carlo simulations~\cite{Clark:2004cp,Clark:2006fx,Clark:2006wp}, describing the partition function
\begin{equation}\label{eq:partition}
    Z(T) = \int DU \hspace{1mm}
    e^{-S_{YM}[U]} \prod_{f=u,d,s} \det(M^{st}_f)^{\frac{1}{4}}
\end{equation}
where the first factor is the exponential of the tree-level Symanzik improved gauge action~\cite{Weisz:1982zw,Curci:1983an}
\begin{equation} \label{symanzik}
    S_{YM}=-\frac{\beta}{3} \sum_{n,\mu\neq\nu}
    \bigg( \frac{5}{6} W_{n;\mu\nu}^{1\times1}
    -\frac{1}{12} W_{n;\mu\nu}^{1\times2} \bigg),
\end{equation}
with $\beta$ denoting the inverse gauge coupling and
$W_{n;\mu\nu}^{1\times1}$ and $W_{n;\mu\nu}^{1\times k}$ representing the real part of the trace of the $1\times k$ rectangle lying on the $\mu\nu$-plane and with first corner on the site $n$. The second factor in the integral of Eq.~\eqref{eq:partition} is the rooted determinant of the stout-improved staggered fermion matrix $M_f^{\rm(st)}$, which reads
\begin{multline}
  M_f^{\rm(st)}(U)_{n,m}
  =     \hat{m}_f\,\delta_{n,m} + \sum_{\nu=1}^{4} \frac{\eta_{n;\nu}}{2}\big[{u_f}_{n;\nu}\,U_{n;\nu}^{(2)}\, \delta_{n,m-\hat{\nu}} \\
  - {u_f}_{n-\hat{\nu};\nu}^*\,U_{n-\hat{\nu};\nu}^{(2)\dagger}\, \delta_{n,m+\hat{\nu}} \big],
\end{multline}
 where $\hat{m}_f = am_f$ is the bare $f$-quark mass, $\eta_{n;\nu} = (-1)^{n_1,n_2, \dots, n_{\nu-1}}$ are the staggered quark phases, $U_{n;\nu}^{(2)}$ are the two-times stout smeared $SU(3)$ link variables, with isotropic smearing parameter $\rho^{\rm stout}=0.15$, and
 \begin{equation}
     {u_f}_{n;\nu} =
 \begin{cases}
     e^{i a^2 q^f B n_x} & {\rm if}\:\: \nu = y, \\
     e^{-i a^2 q^f L_x B n_y} & {\rm if}\:\: \nu = x\:\: \wedge \:\: n_x=L_x,\\
     1 & {\rm otherwise},
 \end{cases}
\end{equation}
are the abelian phases acquired by a $q^f$-charged particle moving along the link in the presence of a constant and uniform magnetic field of intensity $B$ directed along the $z$-axis~\cite{DElia:2012ems,Al-Hashimi:2008quu}. The magnetic field intensity is defined on the smallest electric charge ($|q_f| = 1/3$) as follows
 \begin{equation}
     eB=\frac{6\pi}{a^2L_xL_y}\,b_z\quad {\rm with}\:\: b_z\in Z,
 \end{equation}
where $a$ is the lattice spacing and $L_\mu$ is the lattice extent in the $\mu$ direction in lattice units. Finally, to avoid non-physical cut-off effects, the constraint
\begin{equation}\label{eq:bz_constraint}
    \frac{2b_z}{L_xL_y}\ll1 \, ,
\end{equation}
meaning that $\sqrt{e B}$ is well below the ultraviolet (UV) cut-off, should be respected.

Relying on the analyses conducted in Refs.~\cite{Bali:2011qj,DElia:2021tfb}, the lattice spacing has been considered independent of $eB$, and hence the bare inverse gauge coupling and quark masses have been chosen following Refs.~\cite{Aoki:2009sc,Borsanyi:2010cj,Borsanyi:2013bia}, in order to follow a line of constant physics with physical quark masses. In Table \ref{parameters} we provide a sketch of the simulation parameters which have been used to produce the gauge configuration used in this work.

\section{Computation of the Electrical Conductivity}
\label{sec:BG}

Transport coefficients of a medium related to the electrical conductivity can be extracted from the low-energy behaviour of appropriate current-current spectral functions $\rho_{ij}$ making use of the Kubo formulae
\begin{equation}\label{eq:kubo}
    \frac{\sigma_{ij}}{T} = \frac{1}{2T}\lim_{\omega \rightarrow 0} \frac{\rho_{ij}(\omega)}{\omega},
\end{equation}
where $\sigma$ is the electrical conductivity tensor, $T$ is the temperature and $\rho(\omega)$ the spectral density related to the discrete current-current correlator
\begin{equation}
    C_{ij}(\tau) = \frac{1}{L_s^3} \left\langle J_i(\tau)J_j(0)\right\rangle,
\end{equation}
via the following relation
\begin{equation}
C_{ij}(\tau) = \int_0^\infty \frac{d\omega}{\pi}
    \rho_{ij}(\omega)K(\tau,\omega).
\end{equation}
Here, $J_i(\tau)$ is the $i^{\rm th}$ component of the space-integrated electrical current evaluated at the (Euclidean) time $\tau$.

Thus, the computation of Electrical Conductivity consists of three main steps. First, we measure the current-current correlation functions on the lattice; then, we extract from the correlators the spectral functions using inversion methods; finally, that yields the conductivity via the Kubo formulas. In the following, we discuss these steps in detail starting from the lattice computation of the correlation functions.

\subsection{Staggered current-current correlation functions}
The physical observable to be computed on the lattice is the current-current Euclidean temporal correlator, namely
\begin{equation} \label{corr}
    C_{ij}(\tau) =
    \frac{1}{L_s^3} \langle J_i(\tau) J_j(0) \rangle,
\end{equation}
where $i,j\in\{1,2,3\}$ and $L_s^3$ is the spatial volume. In the framework of stout improved staggered fermions in the presence of a background magnetic field, this current reads
\begin{equation} \label{current}
    J_i(\tau)=\frac{e}{4} \sum_{f} q_f \sum_{\vec{n}}
    \eta_{n,i} \Big( \bar{\chi}_n^f \,\tilde{U}_{n,i}\, \chi_{n+i}^f +
    \bar{\chi}_{n+i}^f\, \Tilde{U}_{n,i}^{\dagger}\, \chi_n^f \Big),
\end{equation}
where we defined a dressed link variable $\tilde{U}_{n,i}={u_f}_{n,i} U_{n,i}^{(2)}$, and  $n = (\tau, \vec{n})$; the sum over $f$ runs over the three quark flavors $u$, $d$ and $s$, and $\bar{\chi}_n^f$ and $\chi_{n}^f$ are the quark fields in the staggered basis. As in Ref.~\cite{Astrakhantsev:2019zkr}, computing Eq.~\eqref{corr} we dismiss the contribution arising from disconnected diagrams.

In order to properly measure the electrical current-current correlator, it is important to notice that, in the staggered fermion formulation, Eq.~\eqref{corr} corresponds to two different operators depending on the timeslice parity
\begin{equation}
s^{e,o}=1-2 \left[\frac{\tau}{a}\text{ mod }2\right].
\end{equation}
Thus, in the Dirac spinor basis $C_{ij}(\tau)$ reads
\begin{equation}\label{eq:stag_corr}
    C_{ij}^{e,o}(\tau) = \sum_{\vec{n}} \big(
    \left\langle A_i(n)\,A_j(0) \right\rangle - s^{e,o}\,
    \left\langle B_i(n)\,B_j(0) \right\rangle \big),
\end{equation}
where
\begin{equation}
    A_i=e\sum_f q_f
    \bar{\psi}^f \gamma_i \psi^f
    \quad\text{and}\quad
    B_i=e\sum_f q_f
    \bar{\psi}^f \gamma_5\gamma_4\gamma_i\psi^f.
\end{equation}
Here, $\psi^f$ is the Dirac spinor of flavour $f$ and $\gamma_{\mu}$ are the $\gamma$-matrices. Note that the operator $A_i$ corresponds to the current of interest, whereas we need to remove the $B_i$ contribution.

If the space is isotropic one can average over the spatial directions $ij$ of the correlation function, but in the presence of an external magnetic field this cannot be done, since the field introduces an anisotropy in the system which breaks the octahedral symmetry of the spatial lattice, leaving a $D_4$ symmetry on the plane orthogonal to the magnetic field direction. Thus, the conductivity in the parallel and perpendicular directions have to be treated separately. In the following, we refer to $C_T$ as the correlation obtained performing the average along the perpendicular directions while we refer to $C_L$ in the case of the longitudinal component.
\subsection{Extraction of Spectral densities}
We have to invert two different relations to extract the even and odd spectral densities, namely
\begin{equation}\label{eq:relationfundbis}
 C_{L,T}^{e,o}(\tau) = \int_0^\infty \frac{d\omega}{\pi}
    \frac{\rho_{L,T}^{e,o}(\omega)}{f(\omega)}K'(\tau,\omega),
\end{equation}
where we redefined the basis function as 
\begin{equation}\label{eq:basisNew}
K'(\tau,\omega) = f(\omega)\frac{\cosh(\omega(\beta/2-\tau))}{\sinh(\omega\beta/2)}.
\end{equation}
Using the Backus-Gilbert method, one constructs the estimator $\bar{\rho}(\omega)$ of the real spectral density $\rho(\omega)$ as a linear combination of the correlation functions computed on the lattice
\begin{equation} \label{eq:Estimator}
     \bar{\rho}_{L,T}^{e,o}(\bar{\omega})=\pi f(\bar{\omega})\sum_i
    g_i(\bar{\omega}) C_{L,T}^{e,o}(\tau_i).
\end{equation}
The advantage of rearranging the relation in the form~\eqref{eq:relationfundbis}, is that if we set $f(\omega)=\omega$ then we can directly extract the ratio $\rho(\omega)/\omega$ in the limit $\omega \rightarrow 0$:
\begin{equation}
\bigg[\frac{\bar{\rho}_{L,T}^{e,o}(\bar{\omega})}{\bar{\omega}}\bigg]_{\bar{\omega}=0} = \pi \sum_i
    g_i(\bar{\omega}) C_{L,T}^{e,o}(\tau_i).  
\end{equation}
This is exactly the quantity that we need to extract the conductivity using Eq.~\eqref{eq:kubo}. The coefficients can be computed by minimizing a suitable functional. In particular, we used the Hansen-Lupot-Tantalo (HLT) method, introduced in Ref.~\cite{Hansen_2019}, already used in numerous phenomenological lattice studies~\cite{Bulava:2021fre,Gambino:2022dvu,ExtendedTwistedMassCollaborationETMC:2022sta,Panero:2023zdr,Frezzotti:2023nun,Evangelista:2023fmt,Alexandrou:2024gpl,Bennett:2024cqv,Frezzotti:2024kqk} and by some of the authors to measure the sphaleron rate~\cite{Bonanno:2023ljc,Bonanno:2023thi,Bonanno:2023xfv,Naviglio:2023fqq}. In Appendix~\ref{Appendix}, we report details about the extrapolation for values of the smearing width $\sigma$ in the limit $\sigma\rightarrow 0$.

\subsection{Computation of conductivity}
The last step consists in the computation of conductivities using the Kubo formulas reported in Eq.~\eqref{eq:kubo}. To cancel the contribution of the correlator $\langle B_i(\tau)B_j(0) \rangle$ in Eq.~\eqref{eq:stag_corr} from the spectral densities, following Ref.~\cite{Astrakhantsev:2019zkr}, we perform the sum $[\rho_{ij}^e/\bar{\omega}]_{\bar{\omega}=0}+[\rho_{ij}^o/\bar{\omega}]_{\bar{\omega}=0}$ so that, in the continuum limit, the spectral function related to the electrical currents correlator is reproduced. Thus, in this formulation the e.m. conductivity $\sigma_{ij}$ is related to the spectral densities $\rho_{ij}^{e,o}(\omega)$ through the modified Kubo formulas
\begin{equation}
    \frac{\sigma_{L,T}}{T} = \frac{1}{2T} 
    \hspace{0.5mm} \lim_{\omega\to0}
    \hspace{0.5mm} \frac{1}{\omega}
    \Big( \rho_{L,T}^e(\omega) + \rho_{L,T}^o(\omega) \Big).
\end{equation}
To reduce ultraviolet (UV) contributions to the spectral function $\rho(\omega)$, as the presence of the magnetic field does not bring further UV divergent contributions, we performed the whole computation on the difference $\Delta C_{eB}^{e,o}= C_{eB}^{e,o}-C_{eB=0}^{e,o}$ in place of the correlator $C_B^{e,o}$. Thus, applying the procedure hereby outlined on $\Delta C_{eB}^{e,o}$, we directly compute, when possible, the change of the conductivity due to the external magnetic field $\Delta \sigma_B= \sigma_B-\sigma_{B=0}$.

\section{Results}
\label{sec:results}

In this Section we show the results obtained for the electrical conductivities respectively in the perpendicular and parallel directions, with respect to the background magnetic field, as a function of the temperature. 
\begin{figure}
    \centering
    \includegraphics[scale=0.35]{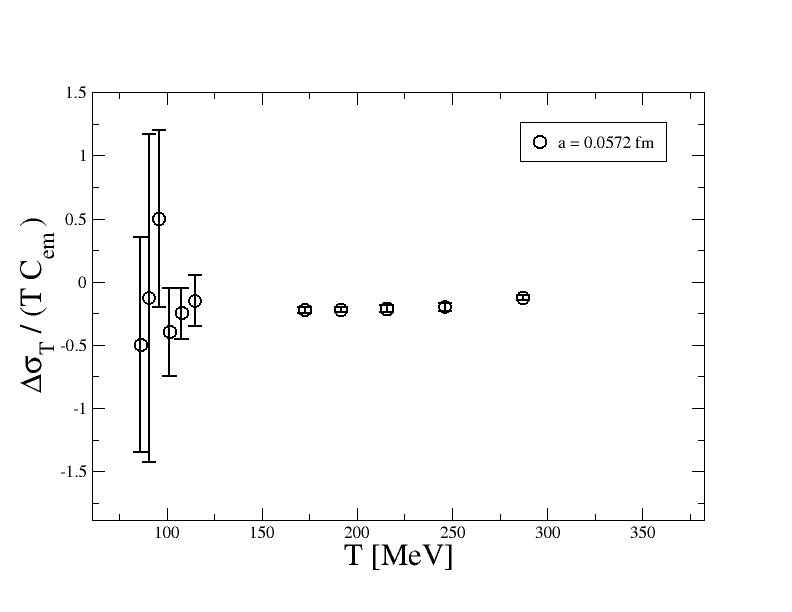}
    \caption{Transverse electrical conductivity due to the magnetic field as a function of the temperature $T=1/L_ta$ for $eB=4\hspace{1mm}$GeV$^2$ and $a=0.0572\hspace{1mm}$fm.} 
    \label{cond_perp}
\end{figure}
\subsection{Conductivity in the transverse direction}
In the presence of a strong magnetic background, the motion in the directions orthogonal to the magnetic field of charged particles is restricted, due to the Lorentz force. Thus, we expect a drop in the electrical conductivity. Such phenomenon, which is often referred to as ``magnetoresistance'', has already been observed in QGP in the computation of the conductivity for magnetic field intensities up to $eB\simeq2.3~{\rm GeV}^2$~\cite{Astrakhantsev:2019zkr}.

In this work, we compute the electrical conductivity variations, $\Delta\sigma_T =\sigma_T(eB=4,9~GeV^2)-\sigma_T(eB=0~GeV^2)$, for a wide range of temperatures. However, we obtain reliable results only for the $eB=4~GeV^2$ case at the finest lattice spacing ($a=0.057~{\rm fm}$), as in all the other cases the noise overwhelms the signal. The results are shown in Figure~\ref{cond_perp}. The magnetic field suppresses the conductivity on the transversal plane as expected, with the magnetoresistance slowly fading out in the high temperature regime. It is reasonable to interpret this as an effect of the increased thermal activity. In the lowest temperature region, below $T\sim100$~MeV, the correlation functions become more noisy. This reflects on the large errors associated to the reconstructed conductivities.

We notice that, in general, substantially
larger errors are always found in the confined phase of the theory, also
for results regarding the conductivity in the longitudinal direction, that we show
in the following subsection; for this particular value of the background magnetic field, the confinement crossover sets in at $T\simeq100$~MeV~\cite{DElia:2021yvk}. These results are not extrapolated to the continuum limit. However, as we will show in the following, for this value of the magnetic field, the discretization effects on the conductivities are substantially under control. Thus, the results obtained in this case on the finest lattice spacing are expected to be faithful to the continuum behaviour.

On the coarsest lattice spacings and in the stronger magnetic field cases, our data are affected by large statistical errors that, together with the weakness of the signal, which is expected to be induced by the suppression of the propagator of charged particles on the plane transverse to $\vec{B}$, make the signal-to-noise ratio of $\sim O(1)$, and hence a meaningful inversion of the Laplace transform does not succeed, making it impossible to produce reliable results.

\begin{figure}
    \centering
    \includegraphics[scale=0.35]{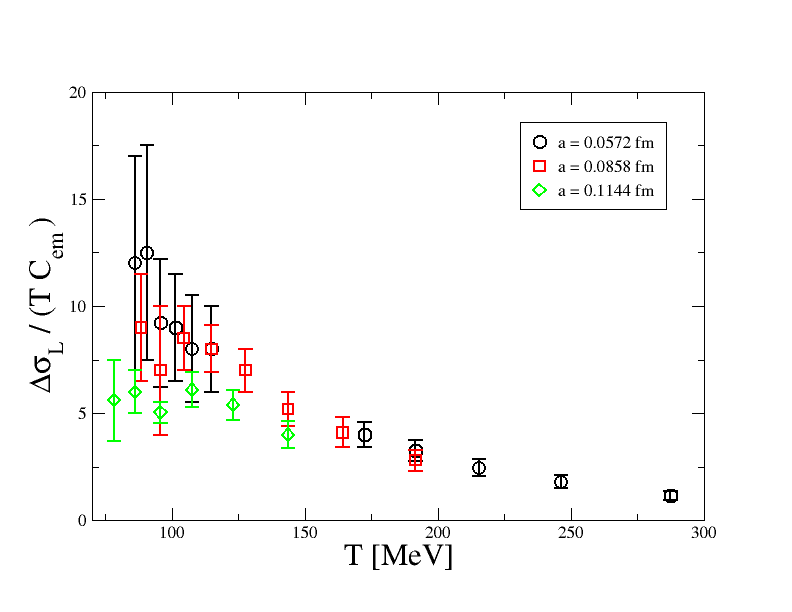}
    \includegraphics[scale=0.35]{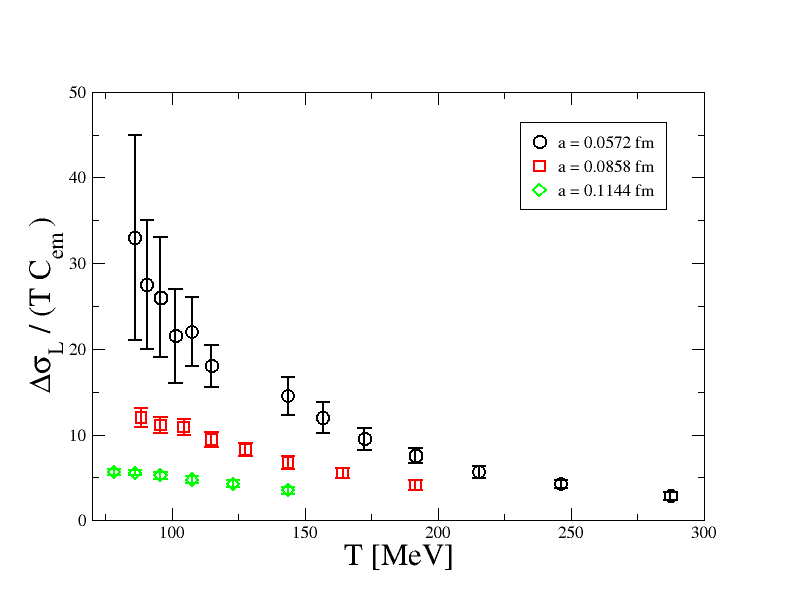}
    \caption{Parallel electrical conductivity due to the magnetic field as a function of $T$ for different magnetic fields and lattice spacings. \textit{Top}: $eB=4\hspace{1mm}$GeV$^2$. \textit{Bottom}: $eB=9\hspace{1mm}$GeV$^2$.} 
    \label{cond_para}
\end{figure}
\subsection{Conductivity in the longitudinal direction}

As already shown in Ref.~\cite{Astrakhantsev:2019zkr}, the magnetic field induces a sizable increase of the conductivity in the longitudinal direction.
This is the reason why for the correlation function in this direction the signal-to-noise ratio of the correlation function is larger and the inversion method is more stable. This allowed us to reliably compute the longitudinal conductivity for the whole range of physical parameters, reported in Table~\ref{parameters}. In Figure~\ref{cond_para} we show its variation as a function of the temperature for both the $eB=4~{\rm GeV}^2$ and the $eB=9~{\rm GeV}^2$ cases. For both the values of the magnetic field, the variation of the conductivity is positive in the whole range of explored parameters, a behavior which is compatible with the CME effect.

In both cases a power-law decreasing behaviour as a function of $T$ can be appreciated: a more detailed discussion on this observation and its physical implications will be presented in the next section.
Regarding the discretization effects, they are generally mild in the 4~GeV$^2$ case, while they are more pronounced for 9~GeV$^2$, in agreement with 
what observed in the determination of other observables, both gluonic and fermionic~\cite{DElia:2021tfb,DElia:2021yvk}. 

\subsection{Transition temperature study}
\begin{figure}
    \centering
    \includegraphics[scale=0.35]{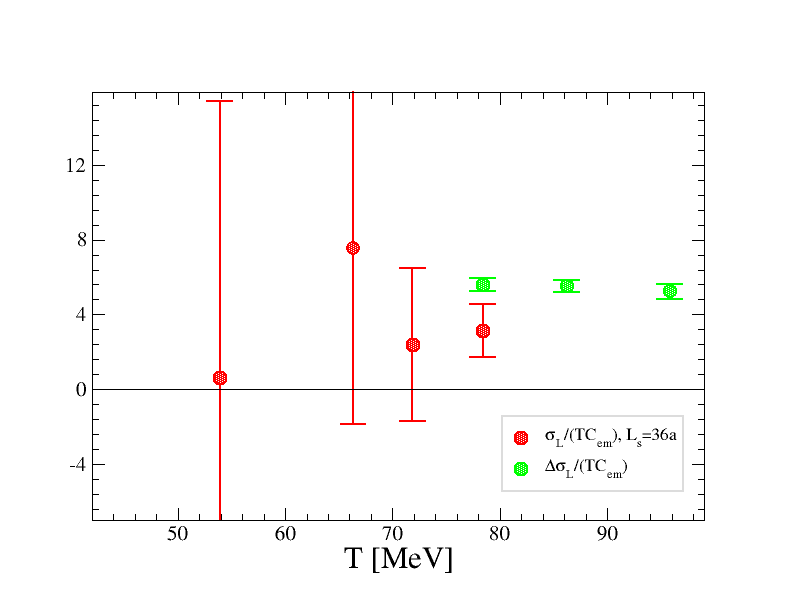}
    \caption{Parallel electrical conductivity as a function of the temperature $T$ for $eB=9\hspace{1mm}$GeV$^2$ and $a=0.1144\hspace{1mm}$fm. The green points are the ones in Figure \ref{cond_para}, so they are $\Delta \sigma_L/TC_{em}$, since they are extracted after the subtraction of the zero magnetic field contribution. {The red points are the unsubtracted measurements performed on lattices with spatial volume $36^3$.}} 
    \label{cond_trans9}
\end{figure}

We performed a preliminary study of the conductivity behaviour across the first order phase transition which is found at $eB=9$~GeV$^2$, which is expected to be clearer with respect to the $eB=4~{\rm GeV}^2$ case. Indeed, in this case the distinction between the two phases is narrow and thus the conductivity is expected to suddenly drop when the temperature falls below its critical value, pointing to the disappearance of visible CME effects, essentially because 
of confinement of quark degrees of freedom.
Such exploration requires to simulate even lower temperatures which, in turn, require larger lattice spacings. For this reason, to carry out this exploratory study, we only used the coarsest lattice spacing, where the transition temperature is $T_c\simeq75$~MeV~\cite{DElia:2021yvk}: we measured the correlators on lattices with space volume of $(36\,a)^3$ and euclidean time extents of $22a,\:24a,\:26a,~32a$, corresponding to temperatures ranging from 54 to 78~MeV. The reconstructed conductivities are displayed in Figure~\ref{cond_trans9}, together with the results on the smaller volume for comparison. {Given the exploratory nature of this measurements, we did not perform the subtraction of the UV contributions on the larger volume correlators. However, we can observe that at the transition temperature, the signal-to-noise ratio drops and the (non subtracted) conductivity is compatible with 0, while it is positive  in the deconfined phase:} this is in agreement with the expectations illustrated above, however, in view of the large discretization and finite volume effects, this is an issue which should be investigated more 
in deep in the future.

\section{Discussion and Conclusions}
\label{sec:discussion}

Fig.~\ref{cond_para} suggests that the parallel conductivity variation in the high temperature regime could follow a power law behaviour in $1/T$ for both  magnetic fields. Such suggestion has been checked by means of a best-fit procedure, using the fit ansatz
\begin{equation}\label{eq:fit_ansatz}
\frac{\Delta\sigma_L}{TC_{em}}=\frac{\alpha}{T^{\beta}},
\end{equation}
to fit data for $T\gtrsim100$~MeV, that is the transition temperature expected for $eB=4$GeV$^2$. Results are shown in Table \ref{fit_test} for the finest lattice. The fit results suggest that errors on the conductivity could be overestimated, causing the reduced chi-squared to be always well below one and the relative errors on the parameter $\alpha$ often of $\mathcal{O}(100\%)$ or larger. However, we find $\beta\simeq2$ with a reasonably good precision for all the lattice spacings and for both values of the magnetic field. Thus, in the following we will continue our analysis assuming this value of 
the exponent, so as to be more predictive on the prefactor $\alpha$: of course our conclusions should be considered in view of this assumption for $\beta$.

\begin{table}[h]
    \centering
    \begin{tabular}{c|c|c|c|c}
        \hline $eB$&T[MeV]$>$&$\alpha$&$\beta$&$\chi^2/{\rm d.o.f.}$\\\hline
        &100&2.6(1.9)$\times10^5$&2.00(14)&3.52/8\\
        9&110&3.7(3.2)$\times10^5$&2.07(17)&2.81/6\\
        &120&1.7(2.2)$\times10^5$&2.35(24)&0.05/5\\\hline
        &100&9.5(9.5)$\times10^4$&1.97(20)&2.11/6\\
        4&110&2.6(3.3)$\times10^5$&2.16(26)&1.01/4\\
        &120&1.1(2.3)$\times10^6$&2.42(39)&0.16/3\\\hline
    \end{tabular}
    \caption{Fit results via the ansatz in Eq.~\eqref{eq:fit_ansatz} for the explored values of the magnetic field and $a=0.0572$fm. For each value of the magnetic field, different temperature ranges have been considered.}
    \label{fit_test}
\end{table}

If we assume that the main contribution to the QGP electrical conductivity comes from the CME conductivity in Eq.~\eqref{eq:approx_sigmaCME}, the behavior of $\Delta\sigma_L$ as a function of $T$ can be imputed exclusively to the relaxation time of the chiral charge, $\tau$. This allows us, under such assumptions, to compute the temperature behaviour of the relaxation time, $\tau(T)$, which on the other hand is an important piece of information 
for the interpretation of phenomenological observations within CME.

\begin{figure}
    \centering
    \includegraphics[scale=0.29]{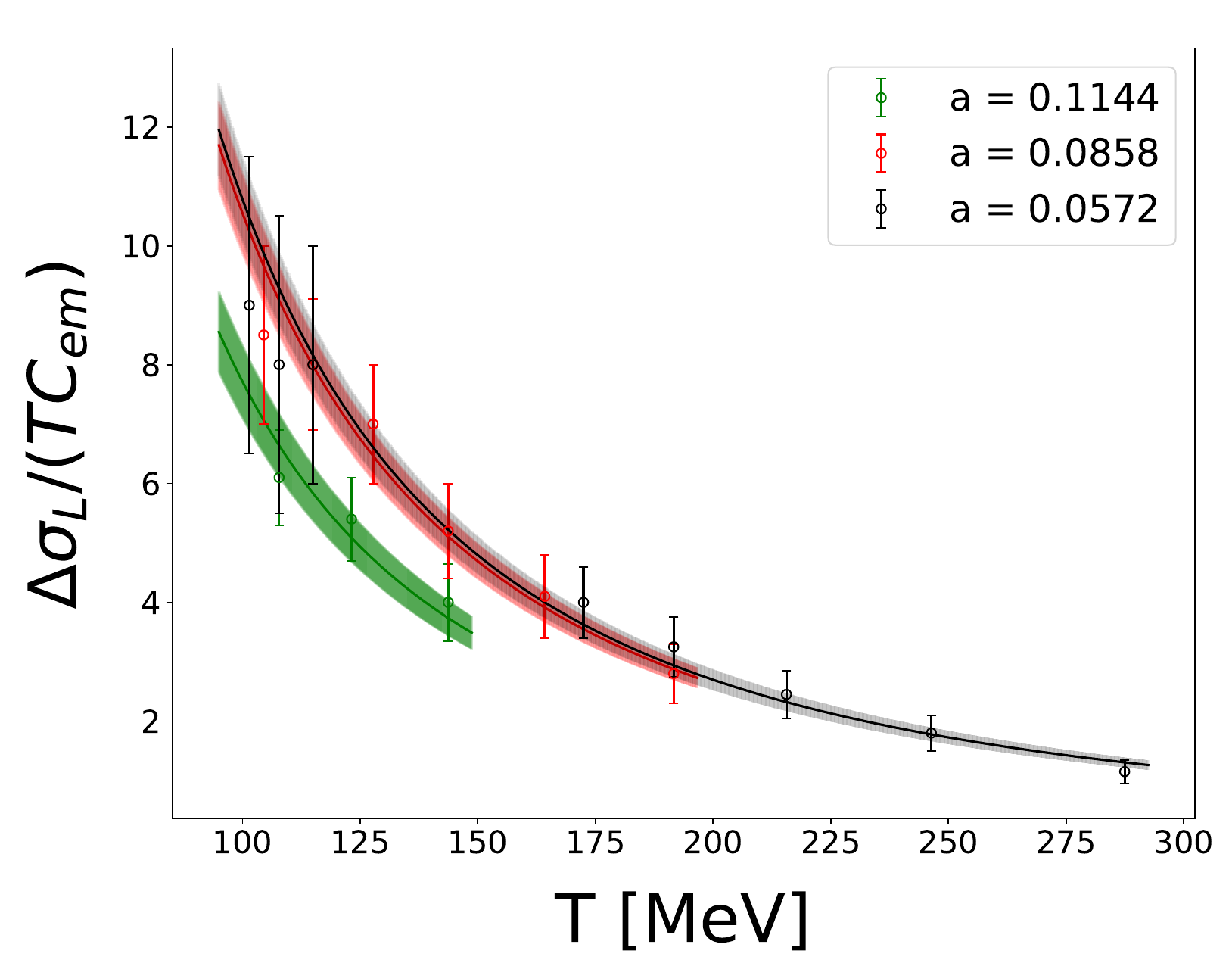}
    \includegraphics[scale=0.29]{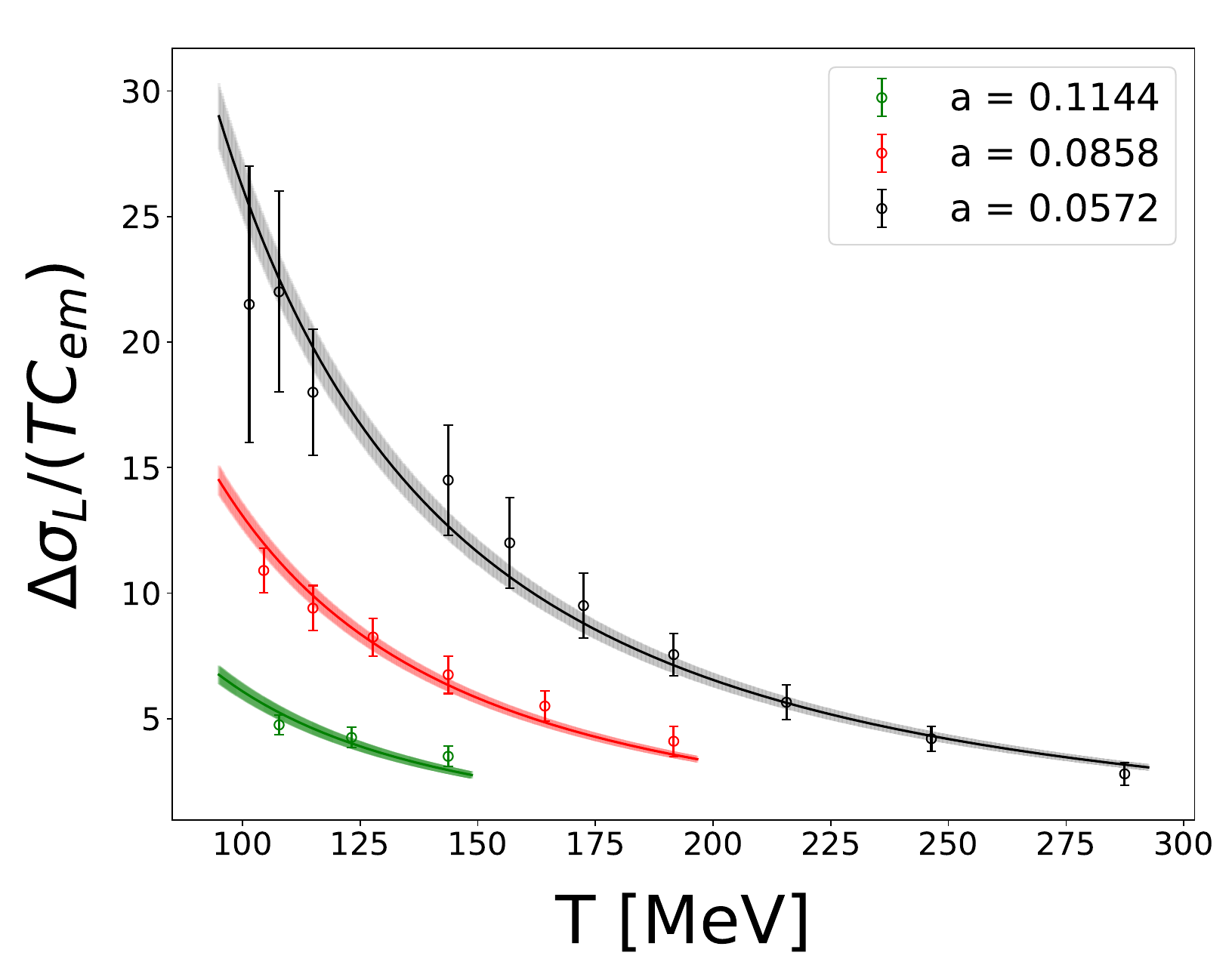}
    \caption{Parallel electrical conductivity due to the magnetic field as a function of $T>120$~MeV for different magnetic fields and lattice spacings. The black band represents the best fit on the $a=0.0572$~fm data sets, the red one is for $a=0.0858$~fm and the green one is for $a=0.1144$~fm, respectively computed for the values in Table \ref{fit_lin} propagating the fit error. \textit{Top}: $eB=4$~GeV$^2$. \textit{Bottom}: $eB=9$~GeV$^2$.} 
    \label{cond_hT}
\end{figure}

Now, since $\sqrt{eB}\gg T$ holds for all of our measures, we can rewrite Eq.~\eqref{eq:approx_sigmaCME} as
\begin{equation} \label{fitlacond}
    \frac{\Delta\sigma_L}{TC_{em}} =
    \frac{C_{em}}{e^2} \frac{N_c}{2\pi^2}
    \frac{eB}{\sum_f |q_f|} \hspace{1mm}
    \frac{\tau}{T}.
\end{equation}
Then, comparing it to Eq.~\eqref{eq:fit_ansatz} with $\beta=2$, one can conclude that, as far as the non-interacting approximation holds, the relaxation time of the chirality-changing processes goes as $\tau\sim1/T$ for $T\gtrsim100$~MeV. Thus, we can provide an estimate for the relaxation time and check the linear behaviour of $\chi(B,T)$  as a function of the magnetic field (see Eq.~\eqref{eq:highB_approx}), so as to give an \emph{a posteriori} proof of the validity of our assumptions. To attain this goal, we make explicit the $T$ dependence of the relaxation time as
\begin{equation}\label{eq:Ctau}
    \tau=C_{\tau}/T,
\end{equation}
and perform the fit on our data using the following improved ansatz
\begin{equation}\label{eq:fit_beta2}
    \frac{\Delta\sigma_L}{TC_{em}}= \frac{3\,eB}{4\pi^2} \,\frac{C_{\tau}}{T^2},
\end{equation}
where the numerical factor has been computed from Eq.~\eqref{fitlacond} using $N_c=3$ and $\displaystyle{\vert q_u\vert = 2\vert q_{d}\vert=2\vert q_s\vert = 2/3}$.  In Figure~\ref{cond_hT} the fit results are displayed together with the measures of the conductivity in the $T \ge 120~{\rm MeV}$ range, and the best fit parameters are listed in Table~\ref{fit_lin} for various temperature ranges.
\begin{table}[h]
    \centering
    \begin{tabular}{c|c|c|c}
        \hline $eB$&$a$[fm]&$C_{\tau},T$[MeV]$>100;110;120$&$\chi^2/{\rm d.o.f.}$\\\hline
        &0.0572&0.122(6);0.123(6);0.125(6)&3.56/9;2.99/7;2.27/6\\
        9&0.0858(2)&0.061(3);0.063(3);0.066(4)&4.11/5;2.06/4;0.68/3\\
        &0.1144(3)&0.028(2);0.031(2);0.031(2)&3.76/2;0.58;0.58\\ 
        \hline
        &0.0572&0.113(7);0.115(8);0.116(8)&2.11/7;1.40/5;1.37/4\\
        4&0.0858(2)&0.110(7);0.113(8);0.114(9)&0.99/5;0.27/4;0.23/3\\
        &0.1144(3)&0.081(6);0.086(9);0.086(9)&0.83/2;0.002;0.002\\ \hline
    \end{tabular}
    \caption{Fit results for all the explored values of the magnetic field and the lattice spacing, using the fit ansatz in Eq.~\eqref{eq:fit_beta2}.}
    \label{fit_lin}
\end{table}

\begin{figure}
    \centering
    \includegraphics[scale=0.3]{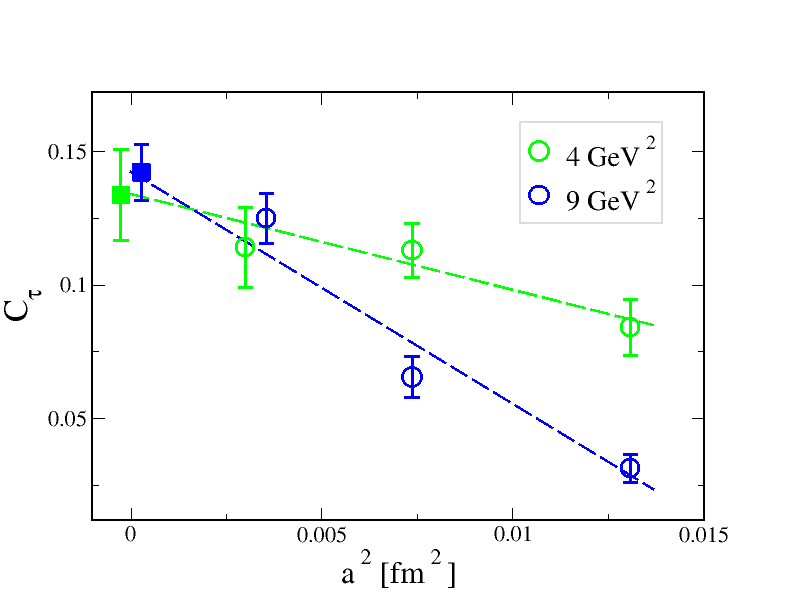}
    \caption{Continuum limit for the fit parameter $C_\tau$. The green and blue points are the results for $C_\tau$ in Table \ref{fit_lin}. The lines are the best linear fit. The filled points are the extrapolated continuum results.} 
    \label{contlim}
\end{figure}

We can see that, in both cases, the one-parameter fits return, in almost all the fit range, a reduced $\chi^2$ lower than $1$. It is also noticeable that the results are mildly dependent on the fit range; nevertheless we use such dependence to obtain an estimate of the systematic error. In Figure~\ref{contlim} results for $C_\tau$ are shown together with the continuum extrapolation. In the $eB=4~{\rm GeV}^2$ case, as already noticed, the discretization effects are very mild, and the linear fit in $a^2$ nicely interpolates data. On the other hand, in the $eB = 9~{\rm GeV}^2$ case, where the finite cut-off effects are much stronger, our best linear fit returns $\chi^2 / d.o.f.\simeq 4.5/1$.

The continuum extrapolations that we obtain for the two values of the magnetic field are:
\begin{equation}\label{result1}
    C_{\tau} (4~{\rm GeV}^2) = 0.134 (17),
\end{equation}
and
\begin{equation}\label{result2}
    C_{\tau} (9~{\rm GeV}^2) = 0.142 (11).
\end{equation}
The compatibility of these results within the errors should corroborate the validity of the non-interacting approximation for the explored values of the magnetic field and temperatures. However, we stress that the study of the finite cut-off effects in the presence of the stronger magnetic field would require finer lattice spacings, as discretization effects on the result in Eq.~\eqref{result2} are not under control.

\begin{figure}
    \centering
    \includegraphics[scale=0.3]{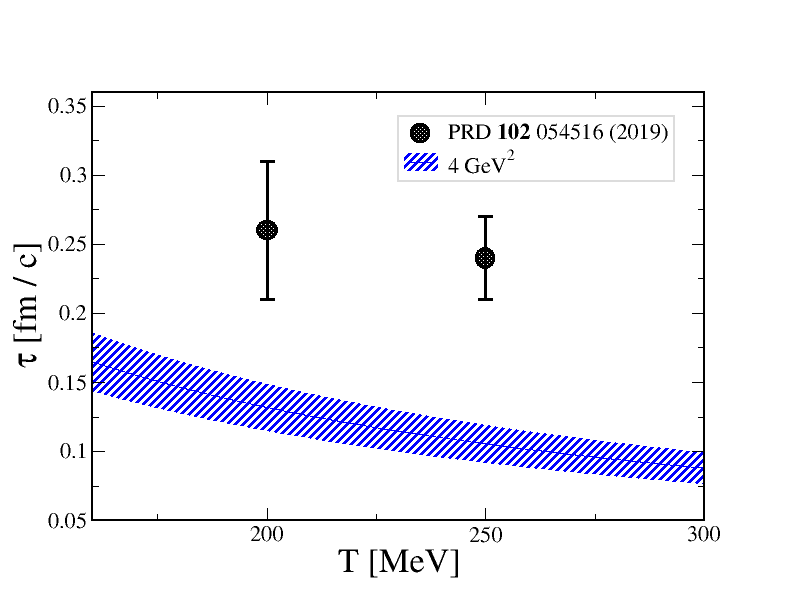}
    \caption{Results for the relaxation time $\tau$ as a function of the temperature $T$.}
    \label{tau_comp}
\end{figure}

We can compare our results with those presented in Ref.~\cite{Astrakhantsev:2019zkr}. Here the authors, relying on the large magnetic field approximation, Eq.~\eqref{eq:highB_approx}, use different values of the magnetic field to fit the parallel conductivity at two fixed value of the temperature, namely $T=200$ and 250~MeV. Then, they extract the relaxation $\tau$ by means of Eq.~\eqref{fitlacond}. 
In Figure~\ref{tau_comp} the black points represent the results reported in Ref.~\cite{Astrakhantsev:2019zkr}, while the colored bands represent our results obtained from Eqs.~\eqref{result1}-\eqref{result2} using Eq.~\eqref{eq:Ctau}. Our results are not in perfect agreement with the findings of the previous work: the discrepancy could be attributed either to the fact that results reported in Ref.~\cite{Astrakhantsev:2019zkr} were obtained for lower values
of $e B$, which are marginally compatible with the large field approximation, or also to the different Laplace inversion method used in that case.

To conclude, we briefly discuss the possible implications of the linear dependence on the inverse temperature observed for $\tau$ at high temperatures. The origin of this dependence lies in the nature of the chirality-flipping processes. Such events can be generated by a finite quark mass $m$ or by sphaleron transitions\footnote{It is possible to exclude the thermal fluctuations in the spin orientation as the origin of such events, as these are suppressed by the $\sqrt{eB}\gg T$ condition}~\cite{hattori}. Indeed, the former breaks explicitly the chiral symmetry, while the latter, consisting in thermal topological excitations, affects the chirality balance through the variation of the topological charge.
Defining $\tau_S$ and $\tau_m$ as the relaxation times associated, respectively, to the sphaleron transitions and the finite quark mass, one can see that~\cite{hattori}
\begin{equation}
    \frac{1}{\tau_S} \sim \frac{\Gamma_S}{\chi T}
    ,\qquad
    \frac{1}{\tau_m} \sim \frac{\alpha_s m^2}{T} \, .
    \label{eq:2tau}
\end{equation}
One may suppose that, under the assumption that quark masses are light enough,
the sphaleron contribution is the dominant one. In Eq.~(\ref{eq:2tau})
$\alpha_s$ is the strong coupling constant, $\Gamma_S$ is the sphaleron rate and $\chi$ is the quantity which appears in the equation of state~(\ref{eq:eos}).

With the above assumption we can justify the observed behaviour of the relaxation time through a dependence on the magnetic field of the sphaleron rate $\Gamma_S$, namely
\begin{equation}
    \Gamma_S \sim eB\hspace{1mm}T^2
\end{equation}
obtained replacing the non interacting approximation for the charge susceptibility $\chi\sim eB$ and the found result $\tau\sim 1/T$ in the above form for the sphaleron inverse relaxation time.
This dependence for the sphaleron rate is in agreement with the result found in~\cite{karzspara} for strong magnetic field, which is properly our case. 
In the future, it would be interesting to check such prediction by direct lattice computations of the sphaleron rate, extending existing results~\cite{Bonanno:2023ljc,Bonanno:2023thi,Bonanno:2023xfv,Naviglio:2023fqq} by including the presence of a magnetic background field.

\begin{acknowledgments}
Numerical simulations have been carried out on the MARCONI100 machine at CINECA, based on the agreement between INFN and CINECA (under project INF20\_npqcd and INF21\_npqcd).  
L.~M.~acknowledges support by the French Centre national de la recherche scientifique (CNRS) under an Emergence@INP 2023 project. 

\end{acknowledgments}

\bibliography{refs}

\appendix\label{Appendix}
\section{}
In this appendix we show how the regularization, represented by the smearing radius $\sigma$, has been removed from the measures of the electrical conductivity.

A key feature of the inversion method introduced in~\cite{Hansen_2019}, and applied in this work, with respect the standard Backus--Gilbert program, is the possibility to choice a suitable kernel function $K(\tau,\omega)$, providing it as an input for the inversion procedure. This allows to make a choice on the smearing amplitude $\sigma$, which is a tunable parameter. Thus, it is possible, to study the dependence of the reconstructed spectral function on $\sigma$.

It is possible to show that, if the target kernel function is even in $\omega/\sigma$, the dependence of the reconstructed spectral density is even in $\sigma$, and hence the absence of odd powers of $\sigma$ in its Taylor expansion:
\begin{equation}\label{eq:sigma_dep}
    \bar{\rho}(\omega;\sigma) = \bar{\rho}(\omega;0) + \frac{\partial^2 \bar{\rho}(\omega;\sigma)}{\partial \sigma^2}  \big\vert_{\sigma=0} \ \sigma^2 + \ \dots \ ,
\end{equation}
where dots represent higher even-power terms. Relying on such property, it is possible to estimate the \emph{zero smearing value} of the reconstructed quantities by means of an extrapolation,

\begin{figure}[t]
    \centering
    \includegraphics[scale=0.28]{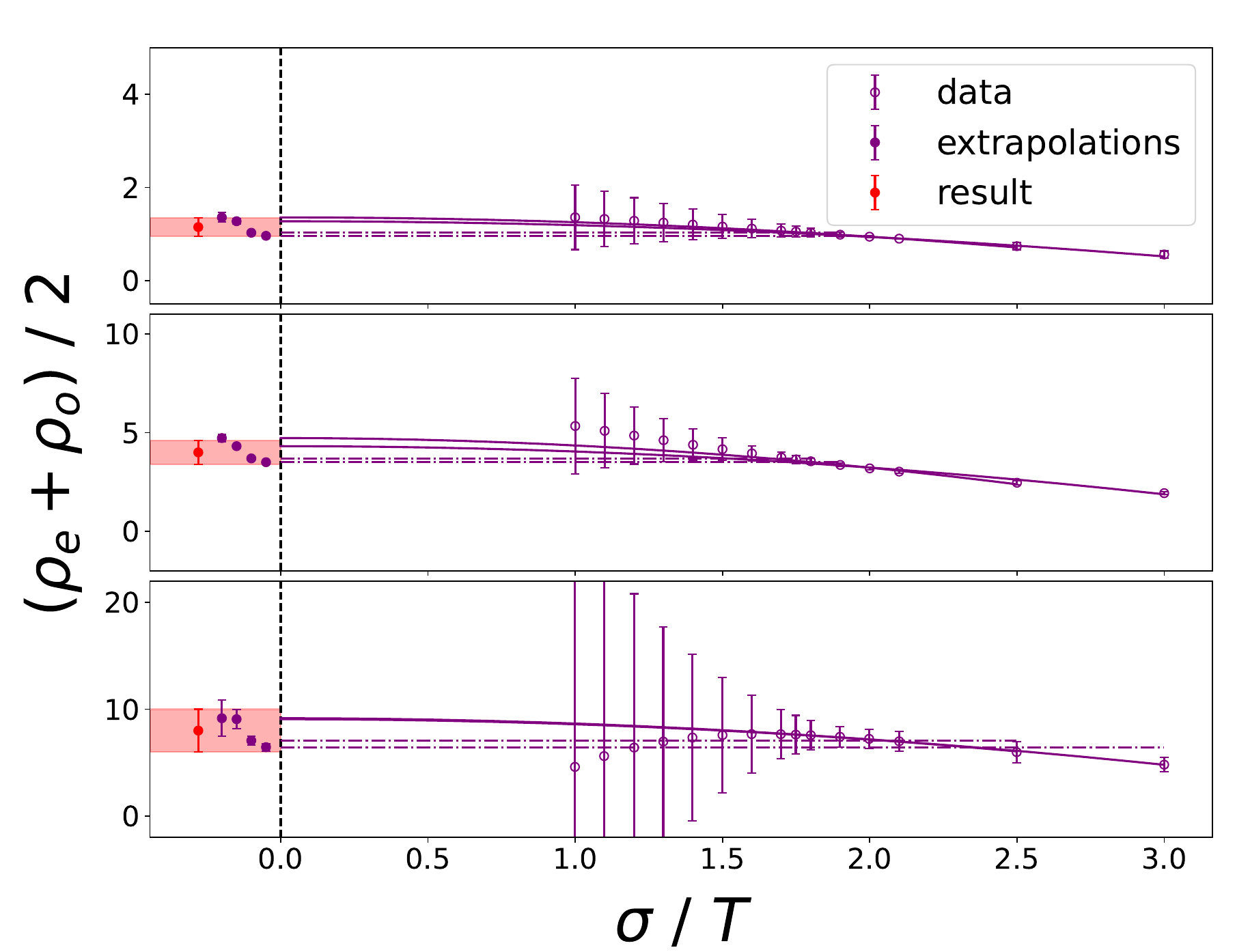}
    \caption{The zero-smearing limit of the electrical conductivity for $eB=4~{\rm GeV}^2$, $a=0.0572$~fm and time extent $N_t=12$, 20, 30 (from top to bottom).}
    \label{fig:app_eB4}
\end{figure}
\begin{figure}[t]
    \centering
    \includegraphics[scale=0.28]{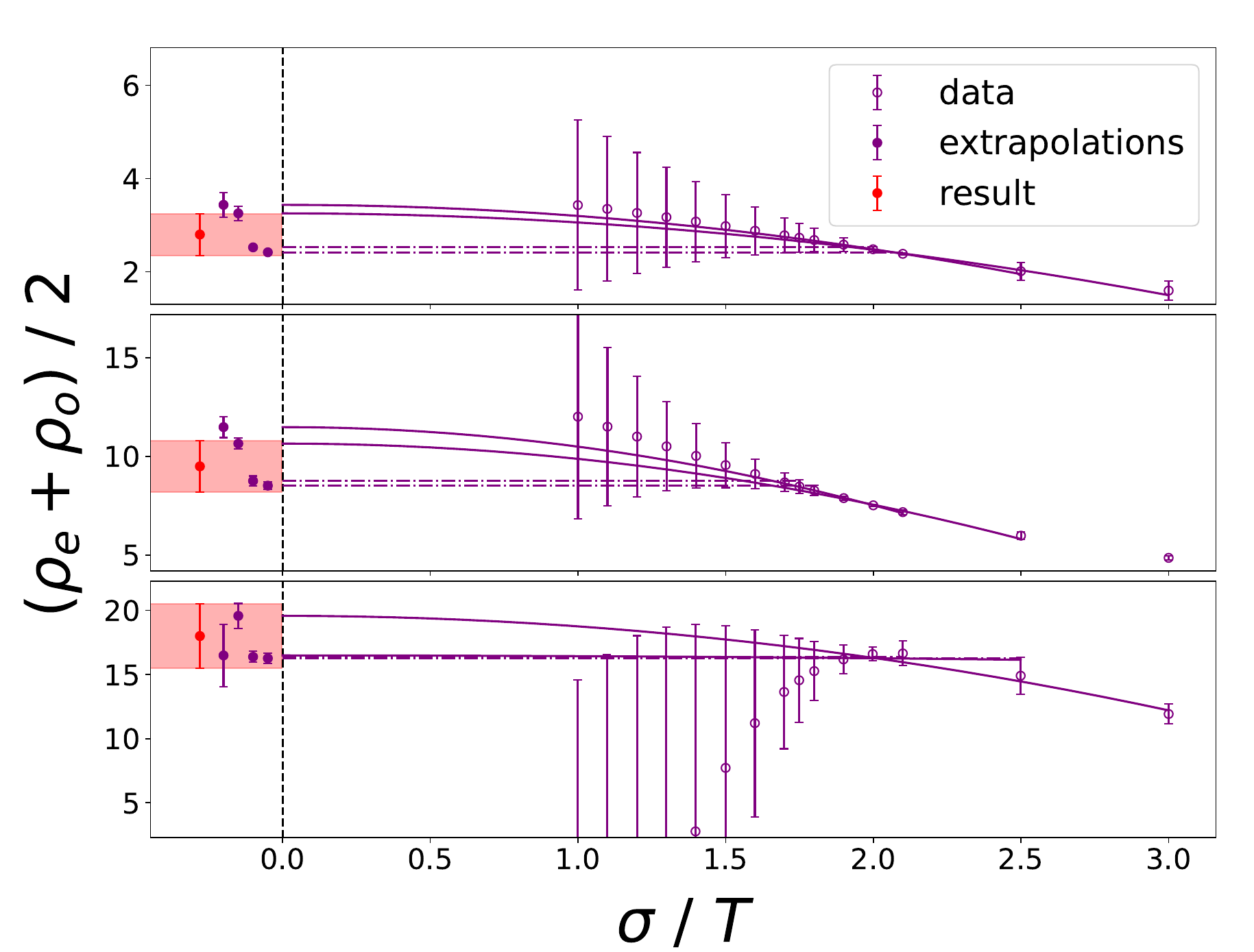}
    \caption{The zero-smearing limit of the electrical conductivity for $eB=9~{\rm GeV}^2$, $a=0.0572$~fm and time extent $N_t=12$, 20, 30 (from top to bottom).}
    \label{fig:app_eB9}
\end{figure}

In Figures~\ref{fig:app_eB4} and~\ref{fig:app_eB9} we show the averages of the reconstructed even and odd spectral densities for several values of $\sigma/T$ in the range $[1,3]$, in a few sample cases: the 12 $a$, 20 $a$ and 30 $a$ temporal extensions, respectively at $eB=4$ and $9~{\rm GeV}^2$ in the finest lattice spacing $a=0.0572$ fm.

The error bars on the data points in the plot consider both statistical noise and systematical effects due to the choice of the parameter $\lambda$ and the difference between the target function and the actual reconstructed kernel, as explained in~\cite{Hansen_2019}. As usual, the signal to noise ratio grows as the smearing radius decreases, due to the increased difficulty in reconstructing the target function. Such effect is particularly evident at the lowest temperature because of the well known exponential decay of the signal to noise ratio at long times in the correlation function.

To estimate systematic effects due to the extrapolation, we used two different $\sigma$-even fit ansatze: constant (dashed lines) and constant plus quadratic term (solid lines); we also considered two different fit ranges to test the fit stability. All the fit shown in these plots present a reduced chi square $\le 1$. The final result and its error are evaluated in order to take conservatively into account all the fluctuations due to the different choices.

\end{document}